\documentclass{article}  
\usepackage{breckenr2005}
\frompage{000} \topage{000}                                              
\def\rightmark{Direct Photon at RHIC-PHENIX}
\def\leftmark{T. Sakaguchi et al.}
 
\title{Direct Photon Measurement at RHIC-PHENIX } 
\authors{
{Takao Sakaguchi for the PHENIX Collaboration %
}\\[2.812mm]
{\normalsize
Brookhaven National Laboratory, \\ 
Physics Department, Upton, NY 11973, U.S.A.\\[0.2ex] 
}
}
 
\abstract{Results on direct photon measurements from the PHENIX experiment at RHIC are presented. The direct photon yields for $p_T>$6\,GeV/$c$ as a function of centrality in Au-Au collisions at $\sqrt{s_{NN}}$=200\,GeV are found to be consistent with NLO pQCD calculation scaled by the number of binary collisions. The results suggest that the photons observed are emitted from the initial stage of hard scattering. Comparisons with several theoretical calculations are also presented.}

 
\begin{document}
 
\maketitle
\setcounter{page}{1}

\section{Introduction}\label{intro}
The Relativistic Heavy Ion Collider (RHIC) at BNL is expected to produce a hot
and dense matter, whose energy density is sufficiently high to induce a phase
transition to the so-called Quark Gluon Plasma (QGP)~\cite{bib1}.
RHIC has been operated for five years since the first collision of Au+Au
in 2000, and produced many intriguing results that have never appeared
in lower center-of-mass energy systems~\cite{bib2}. Especially, the energy
loss of hard-scattered parton in the hot and dense medium that results
in "jet quenching" has been of great interest since its first discovery
by the PHENIX experiment~\cite{bib3}. The hard scattered probe now become
an useful tool to explore system dynamics because its cross-section is large
at RHIC energy, and can be accurately calculated by perturbative QCD (pQCD).
The results obtained so far including jet quenching suggest that
a hot dense medium has been created.

\begin{figure}[htbp]
\centering
\vspace*{-.2cm}
\leavevmode\epsfysize=4.0cm
\epsfbox{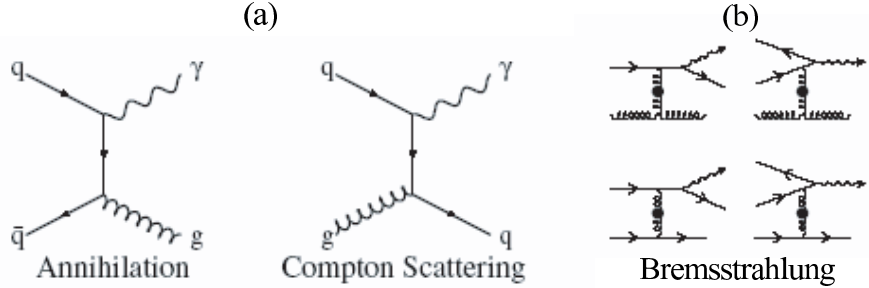}
\vspace*{-.5cm}
\caption[]{Photon production in (a) leading order process, and (b) next-to-leading order process.}
\vspace{-3mm}
\label{figProd}
\end{figure}

\begin{figure}[htbp]
\begin{minipage}{65mm}
\leavevmode\epsfxsize=5.5cm
\epsfbox{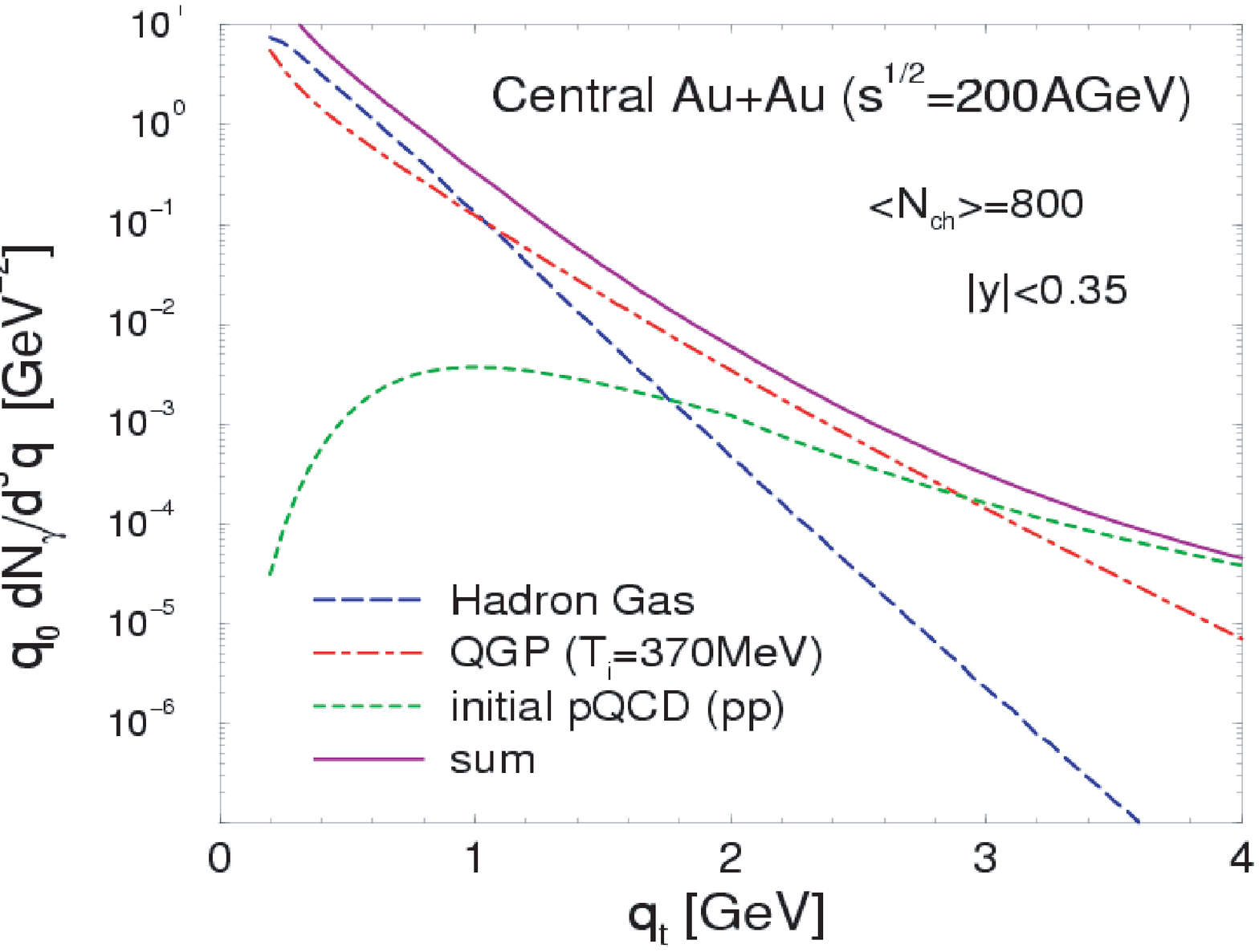}
\vspace*{-.5cm}
\caption[]{Theoretical calculation on various photon contributions for central Au+Au collisions at $\sqrt{s_{NN}}$=200GeV.}
\vspace{-3mm}
\label{fig1}
\end{minipage}
\hspace{1mm}
\begin{minipage}{57mm}
\leavevmode\epsfxsize=5.5cm
\epsfbox{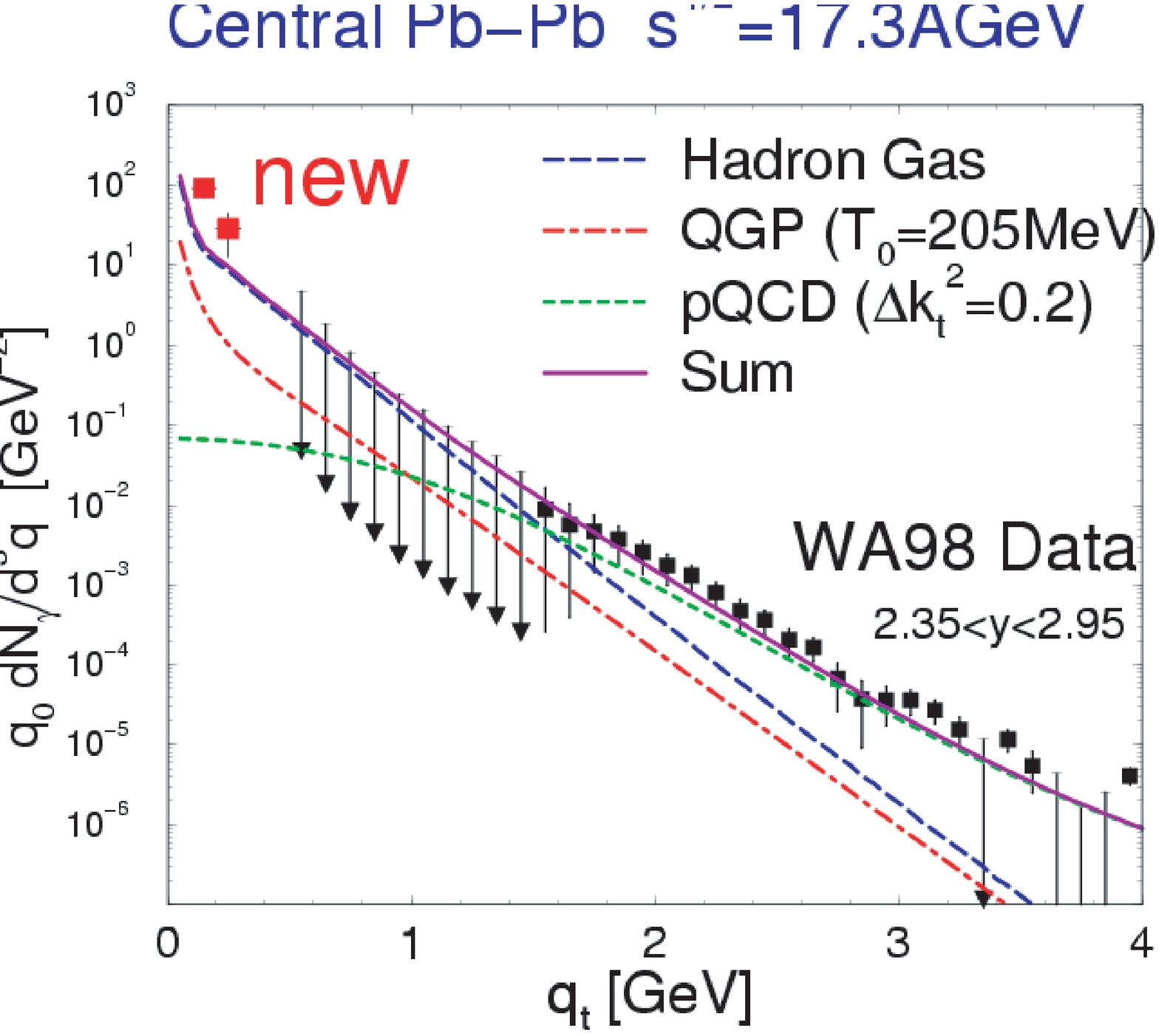}
\vspace*{-.5cm}
\caption{WA98 results and their theoretical interpretation.}
\label{fig4}
\vspace{-3mm}
\end{minipage}
\end{figure}

Photons are excellent electromagnetic probes for extracting the
thermodynamical information of where they are emitted because they do not
interact strongly with medium once produced. Thus, they are expected to
provide hints to answer questions of whether the jet quenching is due
to initial or final state effect, and if the medium turned into the QGP.
The photon production processes are shown in Fig.~\ref{figProd}.
The Compton scattering of quarks and gluons and the annihilation of quarks
and anti-quarks are leading order processes, and the next leading process
is dominated by bremsstrahlung. There is also a prediction of a jet-photon
conversion process, which occurs if QGP is formed,
by a secondary interaction of a hard scattered parton with thermal partons
in the medium~\cite{bib4}. Fig.~\ref{fig1} shows a theoretical prediction
of photons over a $p_T$ range~\cite{bib5}.
The calculation predicts that a photon contribution from the QGP state is
predominant in the $p_T$ range of 1$<p_T<$3\,GeV/$c$. The signal can be seen
after subtracting photons decaying from known hadronic sources. The typical
signal to background ratio is $\sim$10\,\%~\cite{bib21}.
For $p_T>$3\,GeV/$c$, the signal is
dominated by a contribution from initial hard scattering, and $p_T<$1\,GeV,
the signal is from hadron gas through processes of
$\pi\pi(\rho) \rightarrow \gamma \rho(\pi)$, 
$\pi K^* \rightarrow  K \gamma$ and etc..

Direct photon measurements in heavy ion collisions were first carried out
by the WA80 experiment in 200A\,GeV S+Au collisions, where only the upper
limits are obtained~\cite{bib6}. The WA98 experiments observed a significant
excess in the $p_T$ region of 1$<p_T<$3\,GeV/$c$ in 158A\,GeV Pb+Pb
collisions~\cite{bib7,bib7_1}, and a recent theoretical calculation
interpreted it as a consequence of either $k_T$ broadening or high initial
temperature~\cite{bib5_1} as shown in Fig.~\ref{fig4}. Since there is no
direct photon result for p+Pb collisions, one can not evaluate the effect
of $k_T$ broadening in Pb+Pb collisions. The E706 Collaboration at Fermilab
measured direct photons in $\pi$+Be, p+Be at 515 and 800\,GeV/$c$, and showed
that the data deviate from NLO pQCD calculation without including additional
$k_T$ smearing according to the incident particle and target
species~\cite{bib8}. The result suggests that the Cronin-like enhancement can
be explained by the initial $k_T$ broadening. The STAR experiment at RHIC
measured inclusive photons in the $p_T$ region of 0.5$<p_T<$2.5\,GeV/$c$,
and compared it to the possible decay $\gamma$ contributions
from $\pi^0$~\cite{bib9}. Since the systematic error is
order of 40-50\,\% in the measurement, it is not possible to conclude
whether there is a significant excess or not.  

In this paper, the latest results on direct photon measurement from the PHENIX
experiment are shown, and the source of the photon contribution is discussed.

\section{PHENIX detector}\label{techno}  
\begin{figure}[htbp]
\centering
\vspace*{-.2cm}
\leavevmode\epsfysize=5.0cm
\epsfbox{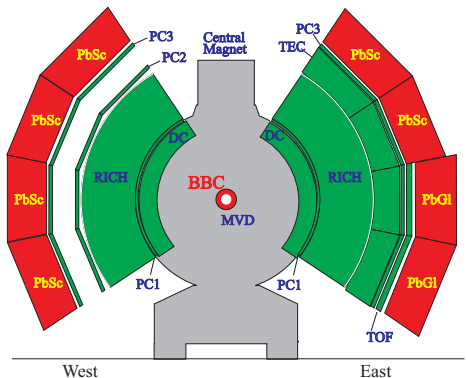}
\vspace*{-.5cm}
\caption{The PHENIX detector.}
\vspace{-2mm}
\label{fig6}
\end{figure}

Figure~\ref{fig6} shows the PHENIX detector at RHIC.
Both central arms (East and West) of the detector include a Multiplicity Vertex
Detector (MVD), Drift Chamber (DC), Pad Chamber (PC), Ring Imaging \v{C}erenkov
Counter (RICH) and Electro-Magnetic Calorimeter (EMCal), and cover a rapidity
range of $|\eta|<$0.35 and a quarter azimuth. The East arm also has a
Time Of Flight (TOF) detector and Time Expansion Chamber (TEC). The detailed
description of the detector is given in the literature~\cite{bib10}.
In this analysis, EMCal was used for measuring the energy of photons.
The EMCal consists of six lead-scintillator sandwich type calorimeters (PbSc)
and two lead-glass homogeneous type calorimeters (PbGl). The DC and PC were
used to track charged particles to estimate charged hadron
contamination in EMCal clusters.

\begin{figure}[htbp]
\centering
\leavevmode\epsfysize=5.0cm
\epsfbox{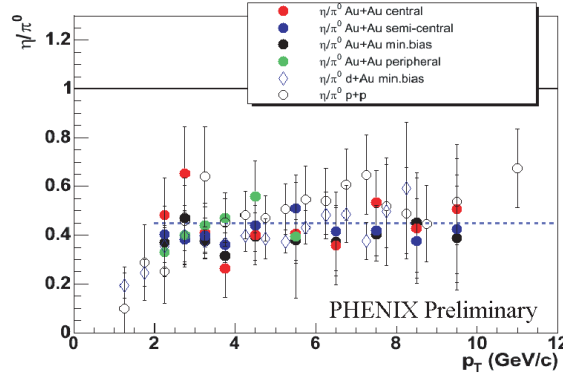}
\vspace*{-.5cm}
\caption{$\eta$ to $\pi^0$ ratio for p+p, d+Au and Au+Au collisions at $\sqrt{s_{NN}}$=200\,GeV. The average value of 0.45$\pm$0.05 is obtained. }
\vspace*{-.5cm}
\label{figEtaPi}
\end{figure}
The analysis of direct photons require determining background photons
decaying from known hadronic sources such as $\pi^0$ and $\eta$. PHENIX has
measured the transverse momentum spectra of $\pi^0$ up to 13\,GeV in Au+Au
collisions~\cite{bib11,bib11_1} and p+p collision~\cite{bib12} at
$\sqrt{s_{NN}}$=200\,GeV. The momentum spectra of $\eta$ and other hadronic
sources were estimated by using a fit to $\pi^0$ $p_T$ spectrum and applying
$m_T$ scaling: $p_T \rightarrow ({p_T}^2-{M_{\pi}}^2+{M_{h}}^2)^{1/2}$.
$\eta$ has been measured but only $\eta/\pi^0$, which is 0.45$\pm$0.05
as seen in Fig.~\ref{figEtaPi} was used to determine the normalization factor
of the $m_T$-scaled spectra relative to $\pi^0$. Using above fits, the number
and spectra of photons decaying from $\pi^0$ and $\eta$ are obtained.

The inclusive photon spectra were reconstructed by applying several photon ID
cuts on the measured cluster energy distributions, and correcting for hadron
contamination and PID efficiency. In order to identify photons, a PID
likelihood function is calculated from several quantities measured by EMCal
such as shower shape or ratio of energies among towers, and apply a threshold.
It resulted in a significant reduction of hadronic clusters in the sample.
Since the ratio of $\gamma$ to $\pi^0$ cancels their common systematic errors,
the excess of the measured photon over the estimated background photon
is evaluated in terms of
$R_{\gamma} \equiv (\gamma/\pi^0)_{measured}/(\gamma/\pi^0)_{estimated}$
(double ratio). The final systematic error on the double ratio is
$\sim$12-16\,\%, and $\sim$15-20\,\% on direct photon spectra in Au+Au
collisions~\cite{bib20}.

\section{Results}
\begin{figure}[htbp]
\begin{minipage}{50mm}
\leavevmode\epsfxsize=3.8cm
\epsfbox{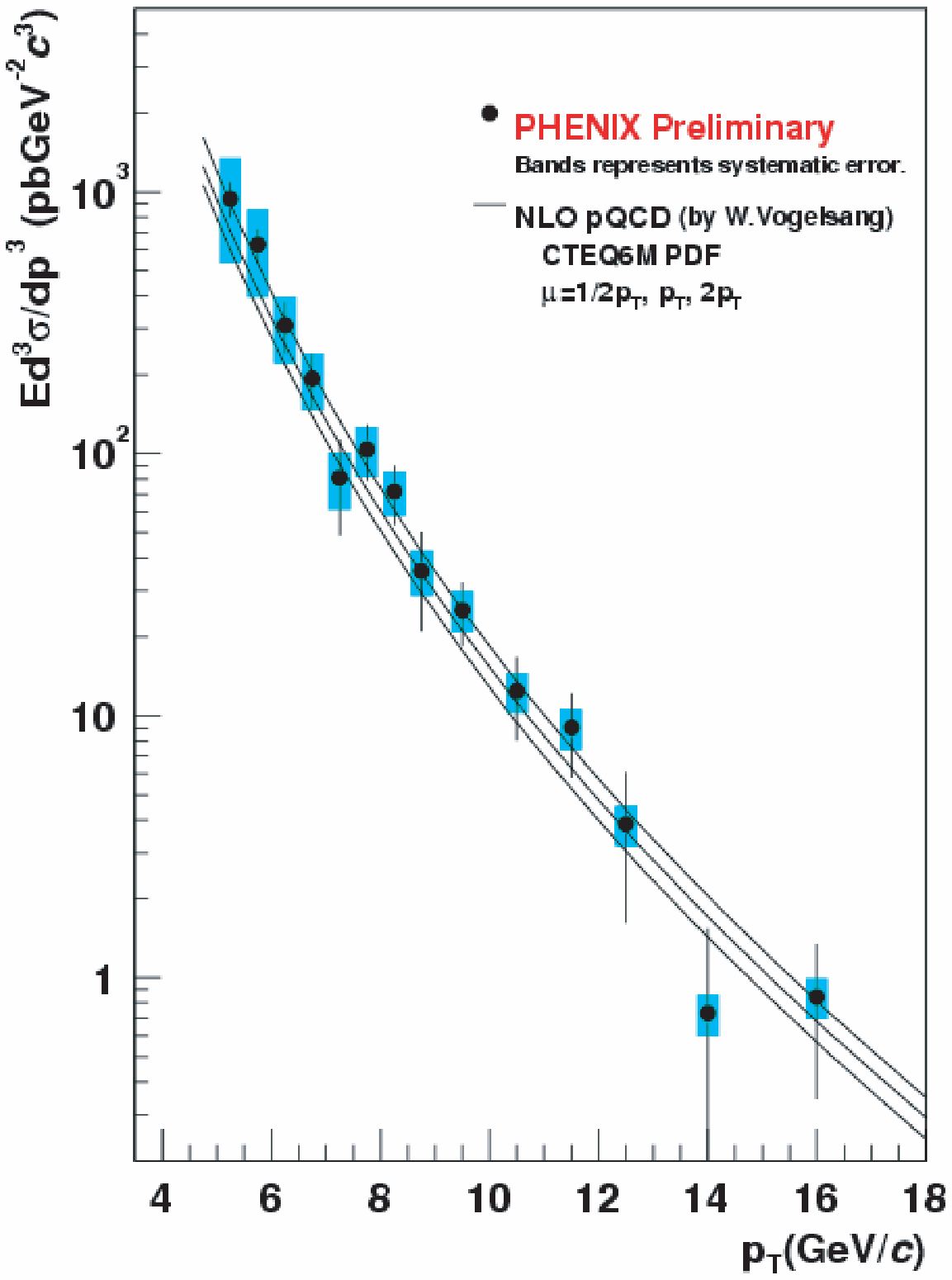}
\vspace*{-.5cm}
\caption{Direct Photon spectrum in p+p collisions at $\sqrt{s}$=200GeV.}
\vspace*{-.4cm}
\label{fig3}
\end{minipage}
\hspace{3mm}
\begin{minipage}{70mm}
\leavevmode\epsfxsize=6.5cm
\epsfbox{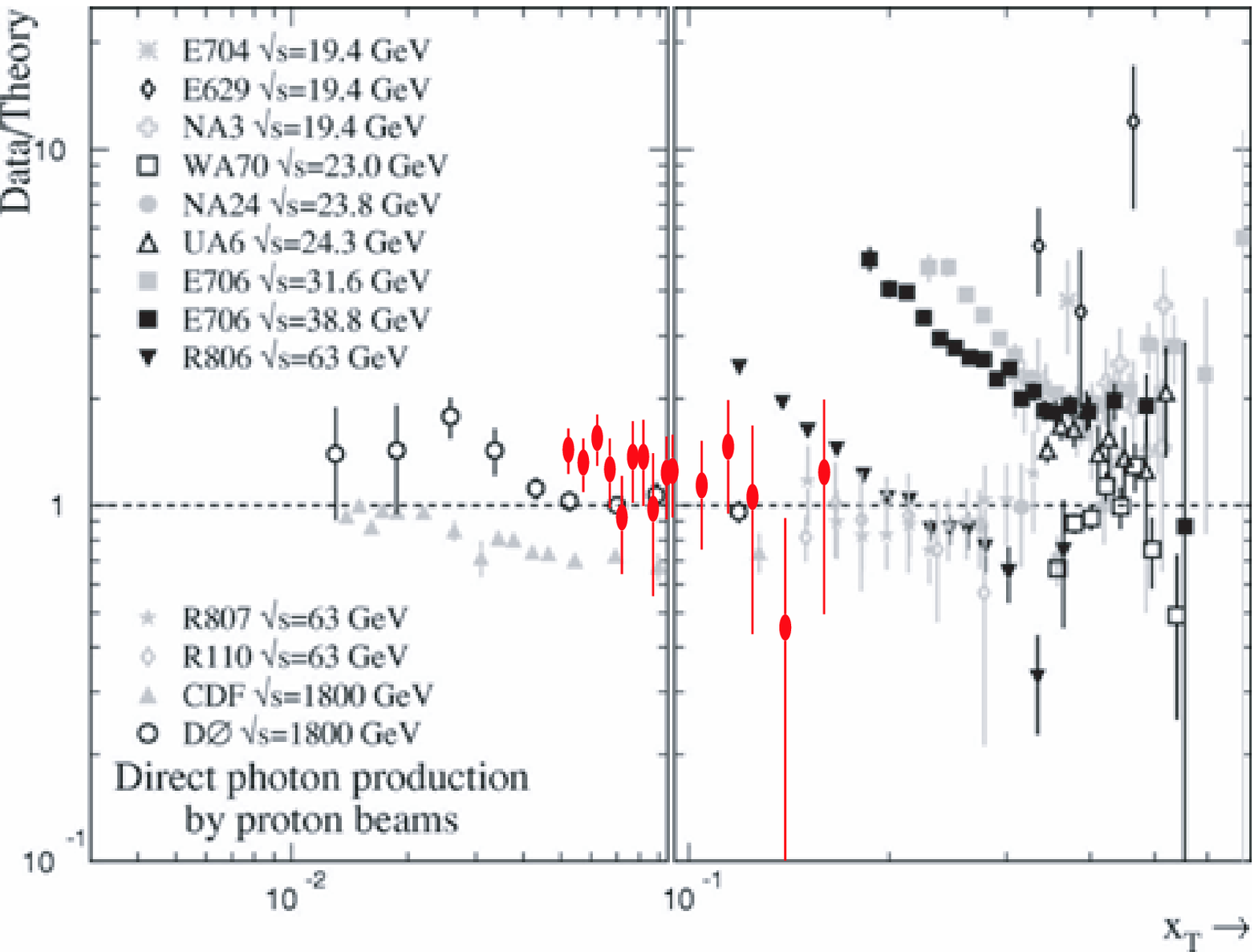}
\vspace*{-.5cm}
\caption{Compilation of comparisons of data to NLO pQCD calculations. PHENIX data are shown as red points.}
\vspace*{-.4cm}
\label{fig13}
\end{minipage}
\end{figure}

\begin{figure}[htbp]
\begin{minipage}{66mm}
\vspace{2mm}
\leavevmode\epsfxsize=6.5cm
\epsfbox{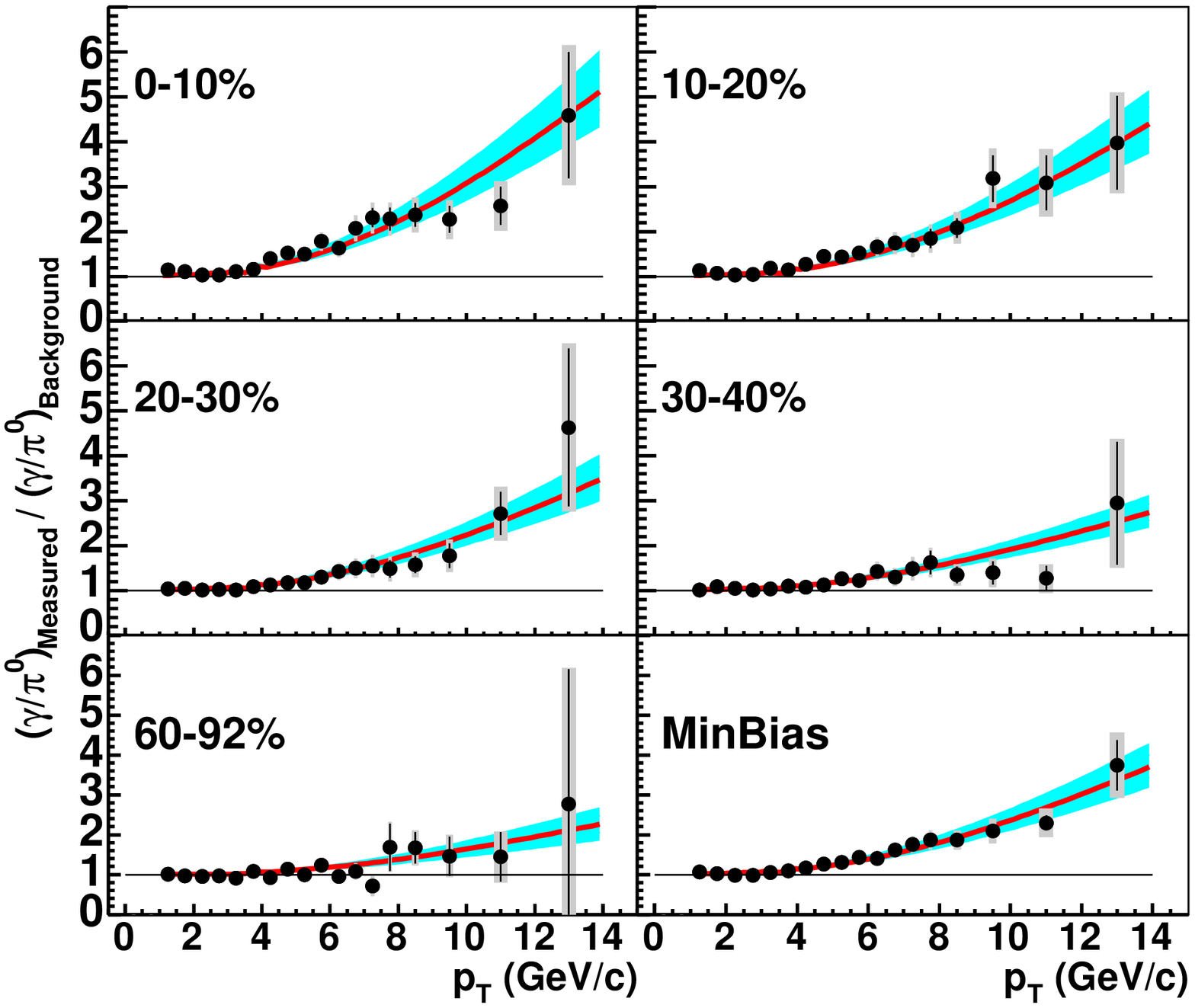}
\vspace*{-.5cm}
\caption{$\gamma/\pi^0$ double ratio in Au+Au collisions.}
\vspace*{-.2cm}
\label{fig7}
\end{minipage}
\hspace{3mm}
\begin{minipage}{54mm}
\vspace{-4mm}
\leavevmode\epsfxsize=5.3cm
\epsfbox{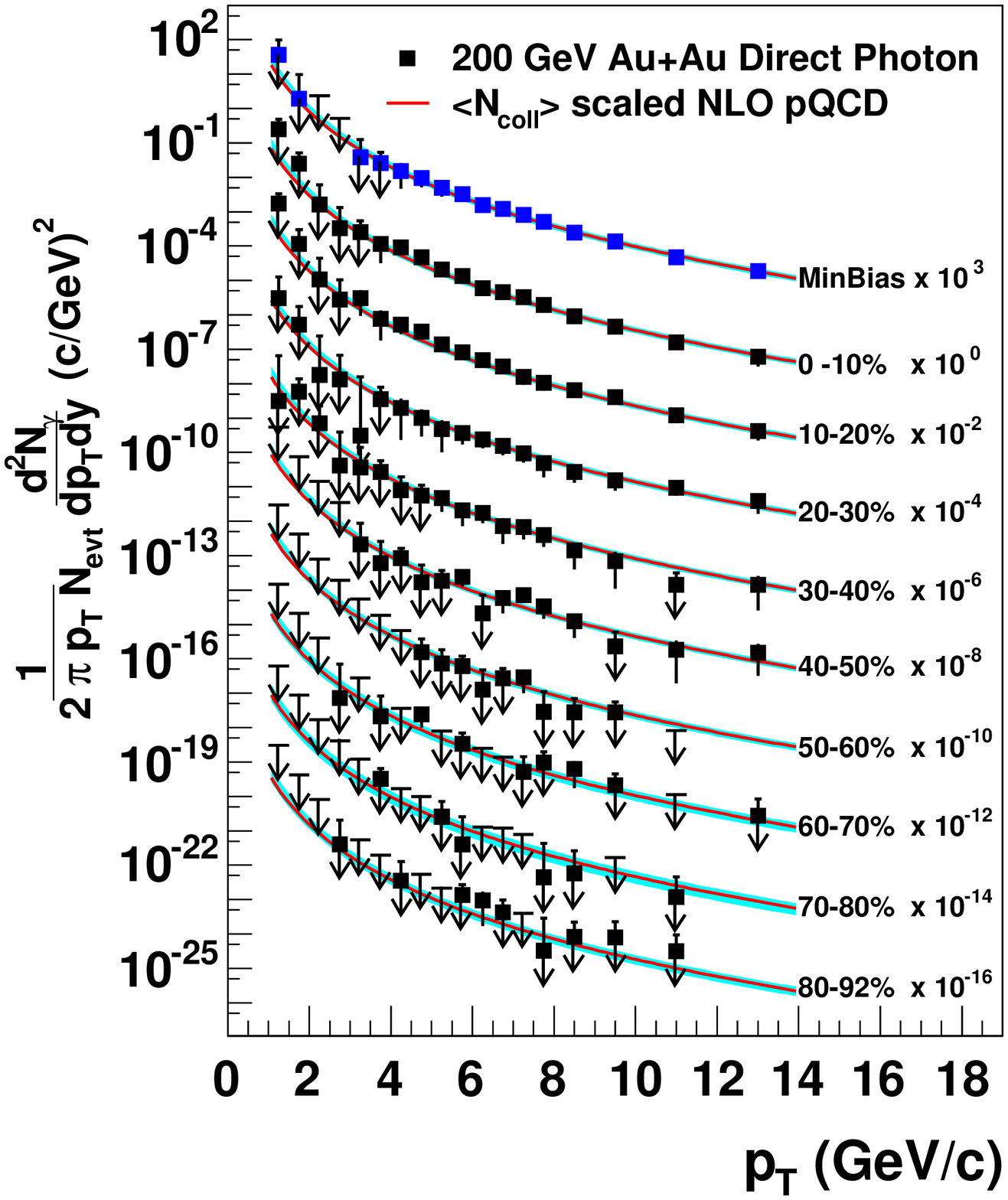}
\vspace*{-.5cm}
\caption{Direct photon spectra in Au+Au collisions over centralities.}
\vspace*{-.2cm}
\label{fig8}
\end{minipage}
\end{figure}

\begin{figure}[htbp]
\centering
\vspace*{-.3cm}
\leavevmode\epsfxsize=9.0cm
\epsfbox{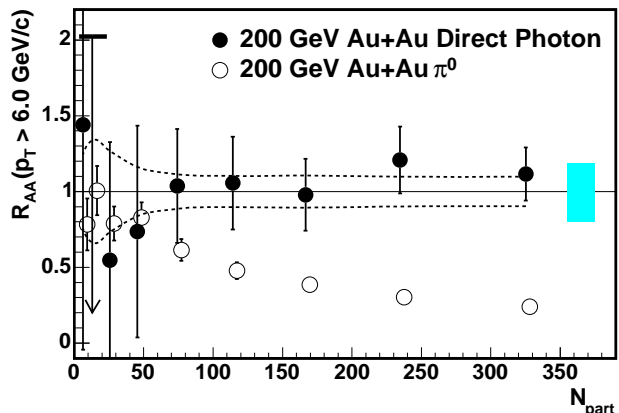}
\vspace*{-.5cm}
\caption{Nuclear Modification Factor ($R_{AA}$) for integrated yield of $p_T>$6\,GeV for $\pi^0$ (open circles) and direct photons (solid circles).}
\vspace*{-.2cm}
\label{fig9}
\end{figure}

Fig.~\ref{fig3} shows the direct photon spectra in p+p collisions together
with the NLO pQCD calculation with three different cut-off scales~\cite{bib14}.
The statistical error is shown as error bars, and the systematic error as
boxes. The data are well described by the calculation.
Fig.~\ref{fig13} shows a compilation of comparisons of data to NLO pQCD 
calculations in p+p and p+A collisions~\cite{bib13}.
As are most of the data points in the figure, the PHENIX p+p data points
shown in red are consistent with pQCD expectation within a factor of 2.
The result shows that the NLO pQCD calculation describes well the yield of
direct photons from the initial hard scattering process in p+p collisions
at $\sqrt{s}$=200\,GeV. Fig.~\ref{fig7} shows the $\gamma/\pi^0$ double
ratio over various centralities in Au+Au collisions together with NLO pQCD
calculation scaled by the number of binary collisions in red lines.
The shaded bands on the data points show the systematic errors, and
the bars are the statistical errors. The shaded bands around the lines show
the uncertainty of the pQCD calculation. The magnitude of the excess at and
above $p_T$ $\sim$4\,GeV/$c$ increases with increasing centrality of collisions,
and is consistent with the calculation. Fig.~\ref{fig8} shows the direct
photon spectra extracted as
$\gamma_{direct}=(1-{R_{\gamma}}^{-1}) \cdot \gamma_{measured}$.
The lines
show the same NLO pQCD calculation scaled by the number of binary collisions. 
It is clearly seen that all nine centralities including minimum bias are
well described by the calculation. Considering the fact that the calculation
in Fig.~\ref{fig7} took the suppression of $\pi^0$ into account which
results in less background photons in higher centralities, and that the
Fig.~\ref{fig8} shows the good agreement between the data and the calculation,
we conclude that the suppression of $\pi^0$ is not attributed to
the initial hard scattering process, while the direct photon is.
Fig.~\ref{fig9} shows nuclear modification factor $R_{AA}$ as a function of
centrality, represented by $N_{part}$, which is defined by:
\[R_{AA}= \frac{(1/{N_{AA}}^{evt}) d^2 N_{AA}/dp_T dy}{<N_{coll}>/\sigma_{pp}^{inel} \cdot d^s \sigma_{pp}/dp_T dy}.\]
The dotted lines show the uncertainty on the number of binary collisions, and
the boxes shows the uncertainty of the NLO pQCD calculation. The sold circle
points are for $\pi^0$~\cite{bib11}, which clearly shows a suppression with
increasing centrality. In case of the direct photons shown in open circles,
the yield follows the number of binary collisions scaling. It once again
proved that the suppression of $\pi^0$ is final state effect.

\section{Discussion}
\begin{figure}[htbp]
\begin{minipage}{65mm}
\vspace*{.1cm}
\leavevmode\epsfxsize=6.5cm
\epsfbox{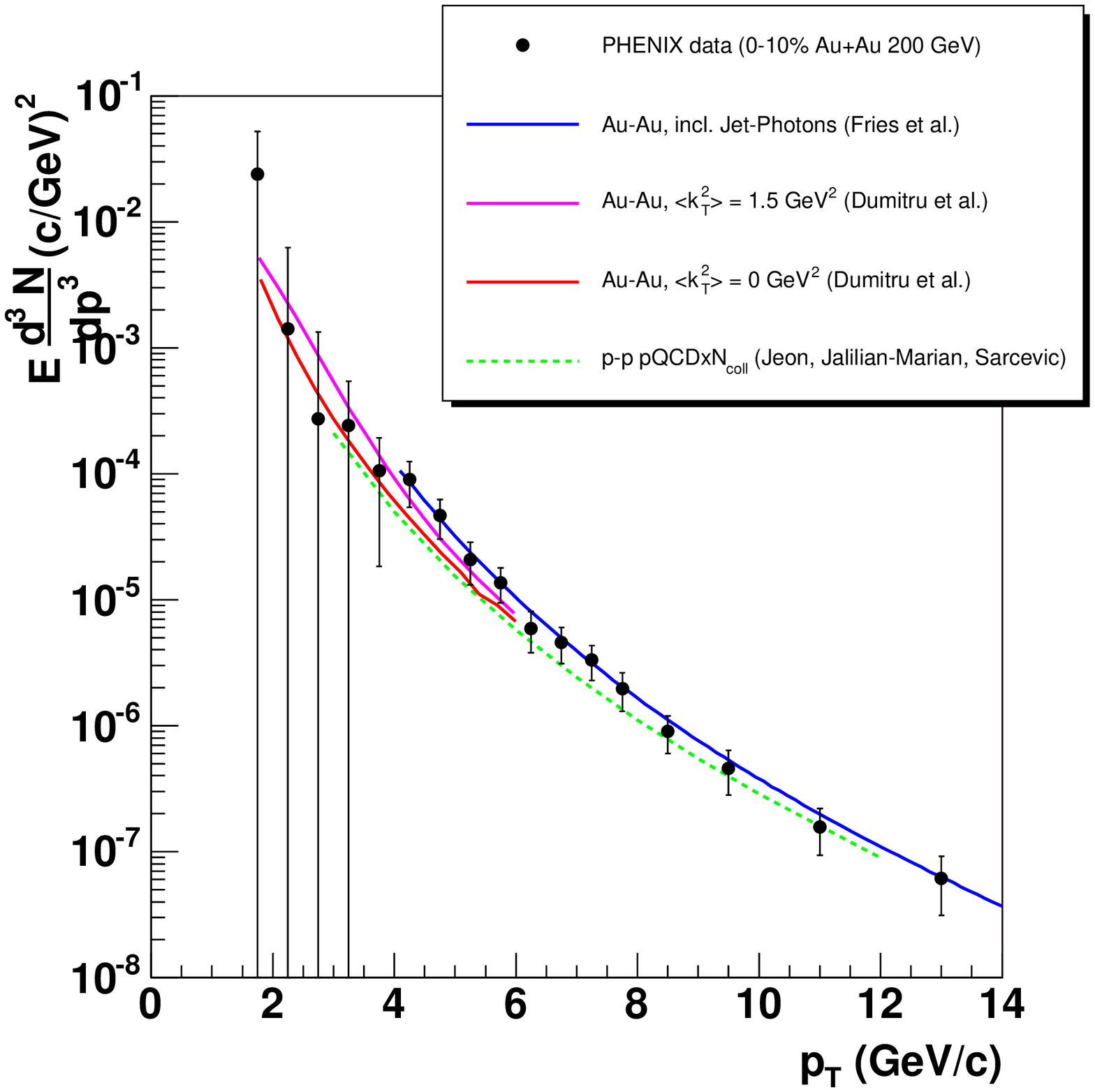}
\vspace*{-.5cm}
\caption{Comparison of 0-10\,\% central PHENIX data with NLO pQCD calculations and jet-photon conversion included model.}
\vspace*{-.5cm}
\label{fig10}
\end{minipage}
\hspace{2mm}
\begin{minipage}{60mm}
\vspace*{-.3cm}
\leavevmode\epsfxsize=6.4cm
\epsfbox{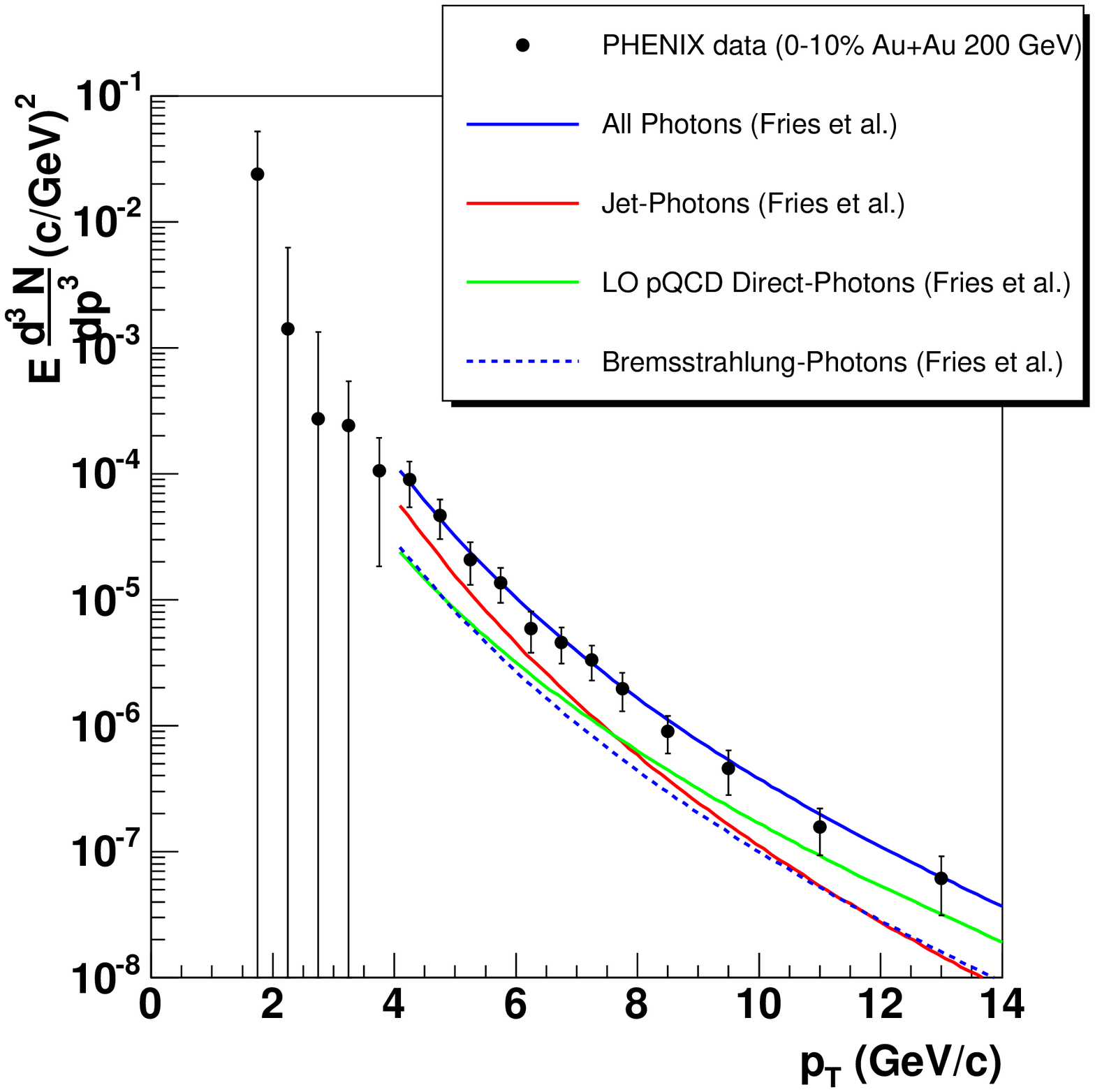}
\vspace*{-.5cm}
\caption{Comparison of the same data with the model with jet-photon conversion to its components.}
\vspace*{-.5cm}
\label{fig11}
\end{minipage}
\end{figure}

\begin{figure}[htbp]
\centering
\leavevmode\epsfxsize=9.0cm
\epsfbox{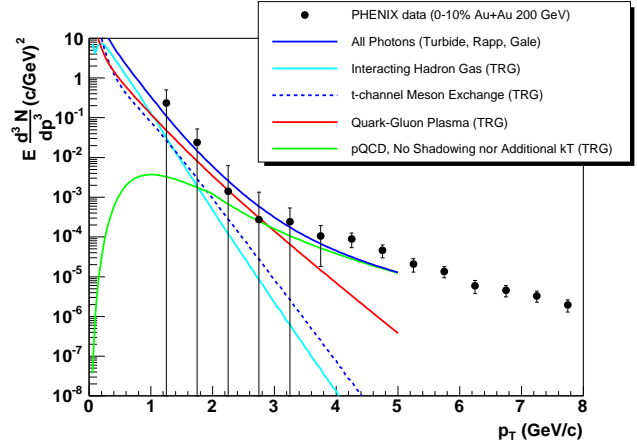}
\vspace*{-.5cm}
\caption{Comparison of the same data with the calculation shown in Fig.~\ref{fig3}.}
\vspace*{-.2cm}
\label{fig12}
\end{figure}

There have been several theoretical calculations on photons from various
contributions. Fig.~\ref{fig10} shows the comparison of 0-10\,\% central
PHENIX data with three NLO pQCD calculations\cite{bib15,bib16}, two of
which have same formulation with different $k_T$ smearing values, and with
the calculation including jet-photon conversion process~\cite{bib4}.
The NLO pQCD calculations with different $k_T$ smearing values show a
difference in the yields below $p_T$ $\sim$ 6\,GeV/$c$, where the data points
have large error and can not distinguish between models. The difference of
yield decreases as $p_T$ increases because of less steep slope at higher $p_T$.
The model including jet-photon conversion describes the data very well
for $p_T>$4\,GeV/$c$. Fig.~\ref{fig11} shows the comparison of the same data
with the model with jet-photon conversion to its components. It can be seen
that the jet-photon conversion process dominates a $p_T$ region of 4-7\,GeV/$c$,
therefore, it should dominate the integrated yields for $p_T>$6\,GeV/$c$.
Accordingly, this yield should follow the number binary collisions
scaling. Thus, the contribution from jet-photon conversion process
should scale with the number of collisions as well. However, the process
assumes the existence of a hot and dense medium or QGP, which is not the case
in peripheral collisions. The model does not give consistent description of
collisions for all centralities.
The data are also compared with calculation on Fig.~\ref{fig3}, and shown
on Fig.~\ref{fig12}.
Within the current error, the calculation is consistent with the data up to
4\,GeV/$c$, and starts deviating from the data above 4\,GeV/$c$. However,
It is possible that the fact that pQCD calculation in the plot does not
include an additional $k_T$ smearing can in part explain the deviation.

\section{Conclusion}
Results on direct photon measurements from the PHENIX experiment at RHIC
are presented. The direct photon yields for $p_T$$>$6\,GeV/$c$ in Au-Au
collisions at $\sqrt{s_{NN}}$=200\,GeV are found to be consistent with
NLO pQCD calculation scaled by the number of binary collisions. The results
suggest that the photons observed are emitted from
the initial stage of hard scattering. Comparisons with several theoretical
calculations are also presented. 

The direct photon analysis in d+Au collisions and high statistics Au+Au
collisions with the goal to reduce systematics error and disentangle
the contributions from QGP state is now ongoing.

\vfill\eject
\end{document}